\documentclass[twocolumn]{aastex631}

\usepackage{textcomp}
\usepackage{subfigure}
\usepackage{appendix}

\usepackage{comment}
\usepackage{rotating}
\usepackage{amsmath}
\usepackage{amssymb}
\usepackage{amsfonts}

\newcommand{\x}{\mbox{$\times$}}
\newcommand{\Msun}{\mbox{$M_{\odot}$}}

\shorttitle{AASTeX v6.3.1}
\shortauthors{Morii et al.}

\begin{document}

\title{The ALMA Survey of 70 $\mu$m Dark High-mass Clumps in Early Stages (ASHES). XI. \\Statistical Study of Early Fragmentation}

\author{Kaho Morii}
\affil{Department of Astronomy, Graduate School of Science, The University of Tokyo, 7-3-1 Hongo, Bunkyo-ku, Tokyo 113-0033, Japan email: kaho.morii@grad.nao.ac.jp}
\affil{National Astronomical Observatory of Japan, National Institutes of Natural Sciences, 2-21-1 Osawa, Mitaka, Tokyo 181-8588, Japan} 

\author{Patricio Sanhueza}
\affil{National Astronomical Observatory of Japan, National Institutes of Natural Sciences, 2-21-1 Osawa, Mitaka, Tokyo 181-8588, Japan} 
\affil{Department of Astronomical Science, SOKENDAI (The Graduate University for Advanced Studies), 2-21-1 Osawa, Mitaka, Tokyo 181-8588, Japan}

\author{Qizhou Zhang}
\affiliation{Center for Astrophysics $|$ Harvard \& Smithsonian, 60 Garden Street, Cambridge, MA 02138, USA}

\author{Fumitaka Nakamura}
\affil{National Astronomical Observatory of Japan, National Institutes of Natural Sciences, 2-21-1 Osawa, Mitaka, Tokyo 181-8588, Japan}
\affil{Department of Astronomical Science, SOKENDAI (The Graduate University for Advanced Studies), 2-21-1 Osawa, Mitaka, Tokyo 181-8588, Japan}
\affil{Department of Astronomy, Graduate School of Science, The University of Tokyo, 7-3-1 Hongo, Bunkyo-ku, Tokyo 113-0033, Japan}

\author{Shanghuo Li}
\affil{Max Planck Institute for Astronomy, K\"{o}nigstuhl 17, D-69117 Heidelberg, Germany}

\author{Giovanni Sabatini}
\affil{INAF, Osservatorio Astrofisico di Arcetri, Largo E. Fermi 5, I-50125, Firenze, Italy}

\author{Fernando A. Olguin}
\affil{Institute of Astronomy and Department of physics, National Tsing Hua University, Hsinchu 30013, Taiwan}

\author{Henrik Beuther}
\affil{Max Planck Institute for Astronomy, Konigstuhl 17, 69117 Heidelberg, Germany} 

\author{Daniel Tafoya}
\affil{Department of Space, Earth and Environment, Chalmers University of Technology, Onsala Space Observatory, 439~92 Onsala, Sweden} 

\author{Natsuko Izumi}
\affil{Academia Sinica Institute of Astronomy and Astrophysics, 11F of AS/NTU Astronomy-Mathematics Building, No.1, Section 4, Roosevelt Road, Taipei 10617, Taiwan}

\author{Ken'ichi Tatematsu}
\affil{National Astronomical Observatory of Japan, National Institutes of Natural Sciences, 2-21-1 Osawa, Mitaka, Tokyo 181-8588, Japan}
\affil{Department of Astronomical Science, SOKENDAI (The Graduate University for Advanced Studies), 2-21-1 Osawa, Mitaka, Tokyo 181-8588, Japan}

\author{Takeshi Sakai}
\affil{Graduate School of Informatics and Engineering, The University of Electro-Communications, Chofu, Tokyo 182-8585, Japan}

\begin{abstract} 
Fragmentation during the early stages of high-mass star formation is crucial for understanding the formation of high-mass clusters. 
We investigated fragmentation within thirty-nine high-mass star-forming clumps as part of the Atacama Large Millimeter/submillimeter Array (ALMA) Survey of 70 $\mu$m Dark High-mass Clumps in Early Stages (ASHES). 
Considering projection effects, we have estimated core separations for 839 cores identified from the continuum emission and found mean values between 0.08 and 0.32 pc within each clump. 
We find compatibility of the observed core separations and masses with the thermal Jeans length and mass, respectively. 
We also present sub-clump structures revealed by the 7 m-array continuum emission. 
Comparison of the Jeans parameters using clump and sub-clump densities with the separation and masses of gravitationally bound cores suggests that they can be explained by clump fragmentation, implying the simultaneous formation of sub-clumps and cores within rather than a step-by-step hierarchical fragmentation. 
The number of cores in each clump positively correlates with the clump surface density and the number expected from the thermal Jeans fragmentation. 
We also find that the higher the fraction of protostellar cores, the larger the dynamic range of the core mass, implying that the cores are growing in mass as the clump evolves.
The ASHES sample exhibits various fragmentation patterns: aligned, scattered, clustered, and sub-clustered. 
Using the $\mathcal{Q}$-parameter, which can help to distinguish between centrally condensed and subclustered spatial core distributions, we finally find that in the early evolutionary stages of high-mass star formation, cores tend to follow a subclustered distribution. 
\end{abstract}


\section{Introduction} \label{sec:intro} 
High-mass stars ($>8$\,\Msun) are mostly formed in clusters \citep[][]{Lada03}. 
Investigating the fragmentation process of molecular clouds provides insights into the formation process of these stellar cluster members through molecular cloud contraction and  has implications to the stellar initial mass function \citep[IMF;][]{Motte22-ALMAIMF}. 
In particular, Infrared dark clouds (IRDCs) are thought to be the best place to study the initial phase of high-mass stellar clusters prior to being significantly disturbed by feedback processes \citep[e.g., outflows, stellar winds, and radiation;][]{Rathborne06, Chambers09, Sanhueza10, Sanhueza12, tan13, Rosen20}. 
Among IRDCs, those that are 70 $\mu$m dark are likely the most pristine, with no evidence of star formation in the IR \citep[e.g.,][]{Sanhueza13, tan13, Guzman15, Sanhueza17, Contreras18}. 
Stars form in gas condensations where gravity dominates over any supporting mechanism such as turbulence, pressure, and magnetic fields. 
To understand what dominates core formation, separations and masses of cores have been compared with those expected from Jeans instabilities \citep{Jeans1902}. 
Considering a uniform, infinite, isothermal medium at rest with a density $\rho$ and pressure $P_0 = \rho c^2_s$, the mean separation of fragments expected from the gravitational collapse, called the thermal Jeans length, is defined as 
\begin{equation}
    \lambda^\mathrm{th}_\mathrm{J} = c_s ( \pi / (G\rho))^{1/2},
    \label{equ:Jlen}
\end{equation} 
where $c_s=( k_\mathrm{B} T / \mu m_\mathrm{H})^{1/2}$ is the isothermal sound speed, $G$ is the gravitational constant, $k_\mathrm{B}$ is the Boltzmann constant, and $m_\mathrm{H}$ is the mass of the hydrogen atom. 
H$_2$ and He govern the thermal velocity dispersion, and the mean molecular weight per free particle $\mu$ can be set to be 2.37 \citep[][]{Kauffmann08}, which is calculated from the cosmic abundance ratios. 
The mass of a sphere associated with the Jeans length is called Jeans mass and is expressed as 
\begin{equation}
    M^\mathrm{th}_\mathrm{J} = \frac{4\pi \rho}{3}\left(\frac{\lambda^\mathrm{th}_\mathrm{J}}{2} \right)^3 = \frac{\pi^{5/2}}{6}\frac{c^3_s}{\sqrt{G^3\rho}}. 
    \label{equ:Jmass}
\end{equation} 
If we include the impact of non-thermal motions as well as the thermal velocity dispersion, the Jeans length and mass are called turbulent Jeans length ($\lambda^\mathrm{tu}_\mathrm{J}$) and mass ($M^\mathrm{tu}_\mathrm{J}$), respectively. 

The recently observed core separations in IRDCs or high-mass star-forming regions are mostly comparable to the thermal Jeans length of clumps rather than to the turbulent Jeans length \citep[e.g., ][ and Ishihara et al. submitted]{Beuther15, Beuther18_CORE, Palau18, Liu19, Sanhueza19, Lu20, Beuther21}. 
For core masses, observations of IRDCs using ALMA revealed that a large portion of cores have low- to intermediate-mass ($\lesssim 30 M_\odot$), which also preferentially exhibit thermal Jeans fragmentation \citep[e.g., ][]{Sanhueza19}, although less sensitive observations at lower resolution with the ALMA 7 m-array or the Submillimeter Array (SMA) have found most massive cores comparable to the turbulent Jeans mass  \citep[$>$ a few\,$M_\odot$;][]{Zhang09, Zhang15}. 
Only a few ALMA studies of more evolved high-mass star-forming regions affected by feedback from massive stars show less effective fragmentation, favoring turbulent fragmentation \citep{Rebolledo20, Jiao23}. 
Some recent studies, on the other hand, propose hierarchical fragmentation \citep[i.e., from clump to sub-clump and sub-clump to cores;][]{Palau18, Pokhrel18,Svoboda19, Rosen20, Zhang21} based on two different spatial resolution observations or double peak of core separation distribution. 
It should be noted that most of these studies are case studies (limited sample) or single-pointing observations (limited spatial area). 
The general trend of core separations and core masses remains unclear, and a statistical study offers the opportunity to conclusively answer the following questions: Are average core properties explained by thermal Jeans fragmentation of clumps? Is clump fragmentation hierarchical? 

We use 839 cores identified from continuum emission of thirty-nine 70 $\mu$m-dark IRDC clumps using the dendrogram algorithm in the ASHES survey \citep{Sanhueza19,Morii23}. 
This is the largest sample of cores embedded in IRDCs and is thought to be the best sample for a statistical study of cores in the very early phase of evolution to date. 
\citet{Morii23} revealed that the majority of the ASHES clumps only host low- to intermediate-mass cores, implying the need for core growth, and that core mass segregation does not clearly appear at such very early stages, although cores are likely segregated in terms of their density. 
This paper is constructed as follows: Section~\ref{sec:obs} describes the observation setups and data reduction process. Section~\ref{sec:result} summarizes the analysis to measure core separations and estimate core masses. We compare observed core properties (separations and masses) with thermal and turbulent Jeans parameters and with Jeans parameters estimated from clump and sub-clump densities, and discuss the hierarchical fragmentation, and diversity found within our sample in Section~\ref{sec:discuss}. Our conclusions are listed in Section~\ref{sec:conclusion}. 

\section{Observations and Data Reduction}
\label{sec:obs}
We have used observations from the ASHES survey, which was carried out with ALMA in Band 6 ($\sim$224 GHz;  $\sim$1.34 mm) through three cycles: Cycle 3 (2015.1.01539.S, PI: P. Sanhueza), Cycle 5 (2017.1.00716.S, PI: P. Sanhueza), and Cycle 6 (2018.1.00192.S, PI: P. Sanhueza). 
The data was taken with the main 12 m array and the Atacama Compact Array (ACA), including both the 7 m array and total power (TP). 
Targets are thirty-nine 70\,$\mu$m-dark IRDC clumps with the potential for high-mass star formation \citep[see Table~1 in][]{Morii23}. 
The whole IRDC clumps were covered by Nyquist-sampled ten-pointing and three-pointing mosaics with the 12 m array and the 7 m array, respectively. 
The mosaicked area corresponds to 0.97 arcmin$^2$ within 20\% power point, equivalent to the effective field of view (FOV) of $\sim$1$'$ per target. 
The detailed observation setups such as on-source time and maximum recoverable scale for all sources are summarized in Table~2 of \citet{Morii23}. 

Our spectral setup includes several molecular lines that have been used in the series of ASHES works: outflow tracers \citep[e.g., CO $J$\,=\,2--1 and SiO $J$\,=\,5--4;][]{Li20, Tafoya21, Morii21}, 
dense gas tracers \citep[e.g., N$_2$D$^+$ $J$\,=\,3--2,  DCN $J$\,=\,3--2, and DCO$^+$ $J$\,=\,3--2;][]{Sakai22}, and shock or warm gas tracers \citep[e.g., H$_2$CO $J_\mathrm{K_a,K_c}$=\,3$_{2,2}$--$2_{2,1}$, H$_2$CO $J_\mathrm{K_a,K_c}$=\,3$_{2,1}$--$2_{2,0}$, CH$_3$OH $J_\mathrm{K}$=\,4$_{2}$--$3_{1}$, and HC$_3$N $J$=24--23;][]{Izumi2-arxiv}. 
The velocity resolution of CO, CH$_3$OH, H$_2$CO, and HC$_3$N 
is $\sim$1.3 km\,s$^{-1}$, and that of other molecules is $\sim$0.17 km\,s$^{-1}$. The detailed description of lines is summarized in \citet[][]{Morii21}. 

Data reduction was carried out using the CASA software package versions 4.5.3, 4.6, 4.7, and 5.4.0 for calibration and 5.4.0 and 5.6.0 for imaging \citep[]{CASA22}. 
Continuum images were produced by averaging line-free channels. 
The effective bandwidth for continuum emission was $\sim$3.7 GHz. 
After subtracting continuum emission, we combined the 12 m array data with the 7 m array data using the CASA task $\mathtt{concat}$, and then they were cleaned together.  
Additionally, we also produced continuum images only from the 7 m array data following the same procedure. 
In this work, we only used TP data of C$^{18}$O ($J$\,=\,2--1) line to estimate the velocity dispersion of the clumps because TP antennas do not provide continuum emission. Some of C$^{18}$O ($J$\,=\,2--1) data has already been presented in the ASHES pilot survey \citep{Sabatini22}. 
We used $\mathtt{TCLEAN}$ with Brigg's robust weighting of 0.5 to the visibilities and an imaging option of MULTISCALE with scales of 0, 5, 15, and 25 times the pixel size, considering the extended components of IRDCs. 
The average 1$\sigma$ root-mean-square (rms) noise level of the combined image is $\sim$0.094 mJy beam$^{-1}$,  with a beam size of $\sim$1$\farcs$2 \citep{Morii23}. 
The root-mean-square (rms) noise level of the 7\,m-array image is summarized in the second column of Table~\ref{tab:cl_Jeans_para}. 

For molecular lines, we used the automatic cleaning algorithm for imaging data cubes, YCLEAN \citep[][]{Contreas_yclean_18, Contreras18} to CLEAN each spectral window with custom-made masks. 
We adopted a Briggs\textquotesingle s robust weighting of 2.0 (natural weighting) to improve the signal-to-noise (S/N) ratio. 
The average synthesized beam size is $\sim 1\farcs 4$. 
The average rms noise levels are $\sim$0.03\,K for cubes with the velocity resolution of 1.3 km\,s$^{-1}$ and $\sim$0.09\,K for cubes with the velocity resolution of 0.17 km\,s$^{-1}$. 
All images have 512 $\times$ 512 pixels with a pixel size of 0$\farcs$2, and all images shown in this paper are the ALMA 12 and 7 m combined, before the primary beam correction, while all measured fluxes are derived from the combined data corrected for the primary beam attenuation. 

\section{Results} 
\label{sec:result}

\subsection{Core Sample}
Using the dendrogram technique \citep{Rosolowsky08}, \citet{Morii23} identified cores using the 1.3 mm continuum images. 
They set a minimum value, $F_\mathrm{min}$, as 2.5$\sigma$, a minimum significance to separate them, $\delta$, as 1.0$\sigma$, and the minimum number of pixels to be contained in the smallest individual structure, $S_\mathrm{min}$, as the half-pixel numbers of the beam area. 
Here, $\sigma$ is a root-mean-square (rms) noise level of the continuum image. 
Since these parameters are optimistic values, they applied the additional constraint to the flux density to exclude suspicious structures. They have excluded cores with a flux density smaller than 3.5$\sigma$. 
Additionally, they eliminated cores at the edge of FOV. 
The total number of identified $\mathtt{leaf}$ structures is 839 from 39 clumps (see an example in Figure~\ref{fig:cont-G24}). 
For the analysis carried out in this work, we have adopted the core masses from \citet{Morii23}. 

\subsection{Classification}
We classified cores into three evolutionary categories and another three categories in terms of their gravitational stability. 
As star-formation signatures, we followed \cite{Sanhueza19} and adopted the detection of molecular outflows traced by CO $J$\,=\,2--1 and SiO $J$\,=\,5--4, and the detection of warm gas tracers (H$_2$CO $J_\mathrm{K_a,K_c}$=\,3$_{2,2}-2_{2,1}$, H$_2$CO $J_\mathrm{K_a,K_c}$=\,3$_{2,1}-2_{2,0}$, CH$_3$OH $J_\mathrm{K}$=\,4$_{2}-3_{1}$, and HC$_3$N $J=24-23$), which all have the upper state energy higher than 45 K. 
We judged cores associated with outflows as outflow cores, and any detection of warm line emission without outflow detections as warm cores. In turn, cores without any detections of outflow and warm gas tracers are classified as prestellar core candidates. This is consistent with the classification of the pilot ASHES survey \citep[][]{Li23}. 
In the end, we have 514 prestellar core candidates, and 325 protostellar cores (222 warm cores and 103 outflow cores). 

To determine the gravitational state of cores, we calculated the virial parameter ($\alpha$), the ratio of the virial mass ($M_{\rm vir}$) to core mass ($M_{\rm core}$). 
The virial mass was calculated as follows:
\begin{equation}
    M_{\mathrm vir} = \frac{5}{a \beta}\frac{\sigma^2_\mathrm{tot} R}{G},
\end{equation}
where $\sigma^2_\mathrm{tot}=\sigma^2_\mathrm{th}+\sigma^2_\mathrm{nt}$ is the velocity dispersion, $R$ is the core radius, $a=(1-b/3)/(1-2b/5)$ is the correction factor for a power-law density profile ($\rho \propto R^{-b}$), and $\beta=(\arcsin e)/e$ is the geometry factor \citep[see][for detailed derivation]{FallFrenck83, Li13}. Here, we adopt a typical density profile index $b=1.6$ \citep[e.g.,][]{Beuther02, Palau14}, and $\beta=1.2$ \citep{Li23}. 
The thermal velocity dispersion and the non-thermal velocity dispersion are given by $\sigma^2_\mathrm{th} = \frac{kT}{\mu m_\mathrm{H}}$ and $\sigma^2_\mathrm{nt} = \sigma^2_\mathrm{obs} - \frac{kT}{m_\mathrm{obs}}$, respectively. 
We assumed that the non-thermal component is independent of the molecular tracer and that $\sigma_\mathrm{obs}$ is the observed velocity dispersion. $m_\mathrm{obs}$ is the molecular weight of the molecule, here 30$m_\mathrm{H}$, corresponding to DCO$^+$ and N$_2$D$^+$. 

The identified cores at early stages of evolution are in dense ($>$10$^5$\, cm$^{-3}$) and cold ($<$20 K) environments where CO depletion and a high-level deuteration occur \citep[e.g.,][]{Caselli02, Sabatini20, Redaelli21, Redaelli22, Sabatini22, Sakai22, Sabatini23}. 
The pilot survey revealed that N$_2$D$^+$ $J=3-2$ and DCO $^+$ $J=3-2$ succeeded in tracing such quiescent dense gas, excellent lines to measure the gas velocity dispersion minimizing the contribution from more diffuse intra-clump gas and gas related to protostellar activity like outflows \citep{Sakai22}. 
We fitted the core-averaged spectrum of the N$_2$D$^+$ and DCO $^+$ taken from the 12 and 7 m arrays combined data with a single Gaussian. 
We judged if the emission is detected with signal-to-noise larger than 3. 
Following the pilot survey \citep{Li22}, to increase the S/N of the weak line emission, the core-averaged spectrum is spectrally smoothed over two native channels, prior to Gaussian fittings, if it shows marginal $\sim$3$\sigma$ confidence in the native spectral resolution. 
Table~\ref{tab:gaussfit} summarize the fitting result such as the peak intensity, the velocity center, and the total velocity dispersion for each line emission. 
Following the discussion in \citet{Li22}, both lines trace the same physical location (dense gas), and if both lines are detected in a core, we generally use $\sigma_{\rm obs}$ measured from DCO$^+$, except for a few cases. 
We adopt the $\sigma_{\rm obs}$ value if either line is detected. 
As a result, we obtained $\sigma_{\rm obs}$ information of 492 cores among 839 cores. 
Following \citet{Li23}, we consider cores with $\alpha <2$ to be gravitationally bound cores, and cores with $\alpha >2$ are unbound, which might be transient objects if we ignore additional support of magnetic fields or external pressure. 
For cores without the detection of DCO$^+$ and N$_2$D$^+$, we classified them as non-detection cores. 
The derived $\alpha$ ranges from 0.06 to $\sim$10 with a median value of 1.2. Among 492 cores, 340 cores are classified as bound cores. 
The detailed results for each core's virial analysis will be presented in a following paper. 

\begin{deluxetable*}{lcccccccc}
\label{tab:gaussfit}
\tabletypesize{\footnotesize}
\tablecaption{Summary of N$_2$D$^+$ and DCO$^+$ Gaussian Fitting Results}
\tablewidth{0pt}
   \tablehead{ Clump Name & Core Name & \multicolumn{3}{c}{N$_2$D$^+$}& & \multicolumn{3}{c}{DCO$^+$} \\ \cline{3-5} \cline{7-9} &  & T$_a$ & $v_{\rm LSR}$ & $\sigma_{\rm obs}$ & & T$_a$ & $v_{\rm LSR}$ & $\sigma_{\rm obs}$ \\ & & (K) & (km\,s$^{-1}$) & (km\,s$^{-1}$)& &(K) & (km\,s$^{-1}$) & (km\,s$^{-1}$)}
   \startdata
   G010.991-00.082&ALMA1&0.28 (0.04)&29.71 (0.06)&0.34 (0.06)&&0.35 (0.03)&29.74 (0.03)&0.29 (0.03)\\
    G010.991-00.082&ALMA2&0.30 (0.04)&29.88 (0.06)&0.38 (0.06)&&...&...&...\\
    G010.991-00.082&ALMA3&0.23 (0.03)&29.98 (0.1)&0.67 (0.1)&&0.37 (0.04)&29.65 (0.06)&0.48 (0.06)\\
    G010.991-00.082&ALMA4&0.18 (0.03)&30.11 (0.09)&0.48 (0.09)&&...&...&...\\
    G010.991-00.082&ALMA5&0.41 (0.07)&29.73 (0.05)&0.27 (0.05)&&0.45 (0.06)&29.85 (0.04)&0.25 (0.04)\\
    G010.991-00.082&ALMA6&...&...&...&&...&...&...\\
    G010.991-00.082&ALMA7&0.44 (0.06)&29.46 (0.05)&0.34 (0.05)&&1.25 (0.06)&29.56 (0.01)&0.21 (0.01)\\
    G010.991-00.082&ALMA8&0.27 (0.04)&30.01 (0.05)&0.3 (0.05)&&0.32 (0.05)&29.81 (0.06)&0.31 (0.06)\\
    G010.991-00.082&ALMA9&0.48 (0.06)&29.5 (0.06)&0.43 (0.06)&&0.52 (0.06)&29.52 (0.05)&0.38 (0.05)\\
    G010.991-00.082&ALMA10&0.26 (0.05)&30.14 (0.04)&0.19 (0.04)&&0.28 (0.04)&30.45 (0.05)&0.28 (0.05)\\
    \enddata
   \tablecomments{The corresponding uncertainty is given in parentheses. Dashes denote no available data. (This table is available in its entirety in machine-readable form.)}
\end{deluxetable*}

\subsection{Core Separation} 
\begin{figure*}
    \centering
    \includegraphics[width=18cm]{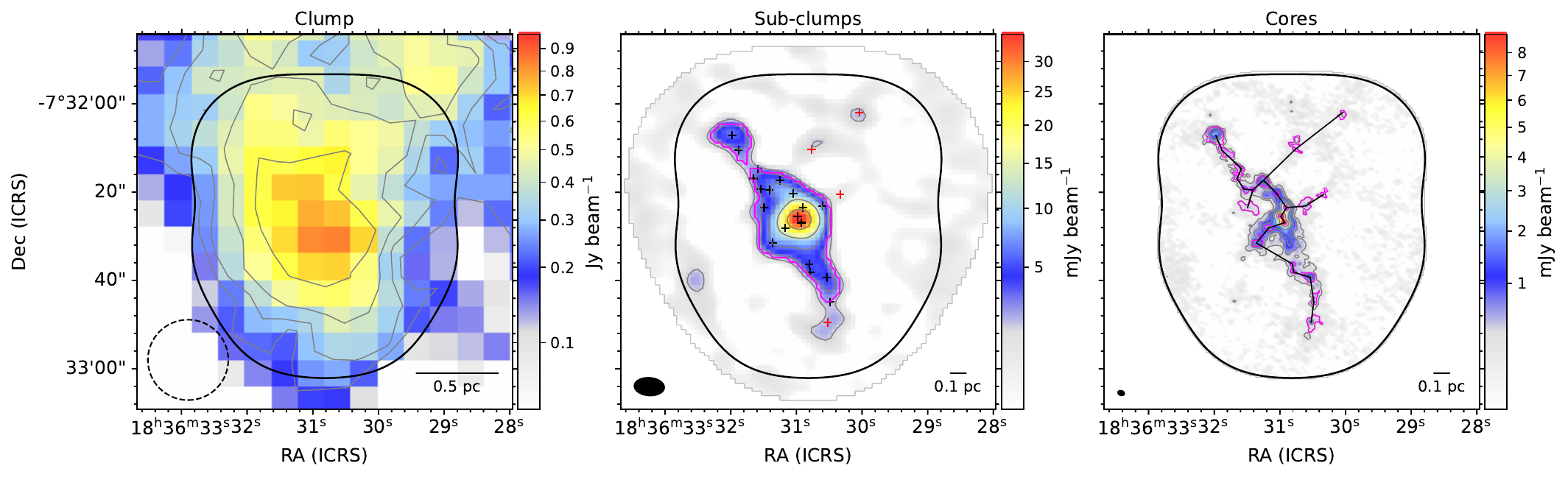}
    \caption{Continuum images for G024.524-00.139 obtained by (left) a single-dish telescope \citep{Schuller09}, (middle) ALMA 7 m array, and (right) ALMA 12 m array and 7 m array. The beam sizes of each image are shown as a black circle or ellipse at the bottom left. 
    (left) Gray-solid contours represent $3 \times 2^\mathrm{n}\sigma$ (n=1, 2, 3, ...), where $\sigma = 71$ mJy\,beam$^{-1}$ is the rms noise level. The black contour represents the FoV of ALMA observations. 
    (middle) Gray-solid contours represent $3 \times 2^\mathrm{n}\sigma$ (n=1, 2, 3, ...) with $\sigma = 0.53$ mJy\,beam$^{-1}$. Magenta thick contour represents leaf structures identified by the dendrogram algorithm. 
    Black and red crosses show core peak positions included in sub-clumps and not-included ones, respectively. The black ellipse in the bottom left corner represents the synthesized beam size. The black line indicates the spatial scale in the bottom right corner. 
    (right) Black segments show the outcome from the minimum spanning tree, which corresponds to the set of straight lines that connect cores in a way that minimizes the sum of the lengths. Gray contour levels are the same with the middle panel but with $\sigma = 0.095$ mJy\,beam$^{-1}$. Magenta thick contour represents leaf structures identified by the dendrogram algorithm.} 
    \label{fig:cont-G24}
\end{figure*}

We used the minimum spanning tree (MST) method developed by \citet[][]{Barrow85} to measure core separation. 
MST connects structures (cores in this case), minimizing the sum of the length and determining a set of straight lines. 
The right panel of Figure~\ref{fig:cont-G24} shows an example of the minimum core separation (edge length determined by MST, hereafter $\delta_{\rm sep}$) as black segments. 
This separation is the projected separation, and the real separation could be equal or longer\footnote{The real separation can be expressed as the projected separation divided by $\cos i$, where $i$ is the inclination angle from the plane of the sky.}. 
Taking the average of the projection effect, we divide the measured separation by a correction factor of $\pi$/4 (see Ishihara et al. submitted for the details). 
The mean $\delta_{\rm sep}$ after correction in each clump ranges from 0.08\,pc to 0.32\,pc, with a median of 0.15\,pc. 

In addition to the minimum core separation, which is the distance between a pair identified by the MST method, we also estimate the nearest separation for each core, which is the distance to the nearest fragments. 
The average of nearest core separation after correction in each clump ranges from 0.07 pc to 0.29 pc, with a median of 0.12\,pc.

\subsection{Sub-clump Identification}
\begin{deluxetable*}{lcccccc}
\label{tab:dendro_subcl}
\tabletypesize{\footnotesize}
\tablecaption{Sub-clump Properties Obtained from Dendrogram}
\tablewidth{0pt}
\tablehead{
\colhead{Clump Name}  & \colhead{Sub-clump Name}  & \colhead{R.A.} & \colhead{Decl.} & \colhead{FWHM$_\mathrm{maj} \times$ FWHM$_\mathrm{min}$}  &\colhead{Peak intensity} & \colhead{Flux Density} \\ 
\colhead{} & \colhead{} & \colhead{(ICRS)} & \colhead{(ICRS)} & \colhead{($''$\,$\times$\,$''$)} &  \colhead{(mJy beam$^{-1}$)} & \colhead{(mJy)}} 
    \startdata
    G010.991--00.082&1&18:10:06.72&-19.27.46.69&12.04$\times$4.26&12.73&23.51\\
    G010.991--00.082&2&18:10:08.20&-19.28.17.69&6.26$\times$3.37&12.67&14.17\\
    G010.991--00.082&3&18:10:07.42&-19.28.02.69&9.66$\times$6.28&12.49&39.73\\
    G014.492--00.139&1&18:17:22.44&-16.25.00.89&6.79$\times$5.72&56.71&110.46\\
    G014.492--00.139&2&18:17:21.54&-16.25.02.89&5.42$\times$3.22&44.68&41.71\\
    G014.492--00.139&3&18:17:22.23&-16.25.30.89&5.88$\times$3.47&24.25&26.03\\\enddata
\tablenotetext{}{(This table is available in its entirety in machine-readable form.)}
\end{deluxetable*} 

Inside clumps, cores are not always uniformly distributed but consist of sub-clusters.  
To study the fragmentation properties of such sub-clusters, we applied the dendrogram technique for the continuum images produced by data only taken by the 7 m-array. 
We used the same parameters as those used for the core identification but with the rms noise level measured in the 7 m-array continuum images (Table~\ref{tab:cl_Jeans_para}). 
For example, the middle panel of Figure~\ref{fig:cont-G24} shows the 7 m-array continuum image for G024.524-00.139, and magenta contours represent the identified leaf structures. 
We define these leaves as sub-clumps. From all clumps, we identified from one to seven sub-clumps per clump, 135 sub-clumps in total.  
Table~\ref{tab:dendro_subcl} gives peak position, beam-convolved size (major and minor full-width half maximum; FWHM), peak intensity, and flux density of sub-clumps identified by the dendrogram algorithm (the properties for all sub-clumps are summarized in a machine-readable table). 

Masses of sub-clumps are estimated from the flux density of 1.3\,mm continuum emission assuming optically thin conditions by 
\begin{equation}
    M_\mathrm{core} = \mathbb{R}\frac{d^2 F_\nu}{\kappa_\nu B_\nu (T_\mathrm{dust})},
    \label{equ:mass}
\end{equation} 
where $\mathbb{R} =$ 100 is the gas-to-dust mass ratio, $\kappa_\nu$ is the dust absorption coefficient, $d$ is the heliocentric distance associated with each ASHES clump, and $B_\nu$ is the Planck function for a dust temperature $T_\mathrm{dust}$. 
We adopt $\kappa_\nu$ of 0.9 cm$^2$\,g$^{-1}$ from the dust coagulation model of the MRN \citep{Mathis77} distribution with thin ice mantles at a number density of 10$^6$\,cm$^{-3}$ computed by \citet{Ossenkopf94}. 
We used the clump dust temperature listed in \citet{Morii23}. 
As the ASHES clumps are in their early stages, the free-free contamination especially at 1.3 mm is negligible. Here, we assume that the continuum emission comes only from dust emission. 

The surface density, $\Sigma$, and the molecular volume density, $n(\mathrm{H_2})$, were estimated assuming a uniform spherical density distribution as:  $\Sigma=M_\mathrm{sub-cl}/\pi R^2_\mathrm{sub-cl}$ and $n(\mathrm{H_2})=M_\mathrm{sub-cl} / \Bar{m}_\mathrm{H_2}(4\pi R^3_\mathrm{sub-cl}/3)$, where $R_\mathrm{sub-cl}$ is half of the geometric mean of the FWHM (Table~\ref{tab:dendro_subcl}). 
In this paper, we defined cores overlapped with the sub-clump as members of the sub-clump (black crosses in the middle panel of Figure~\ref{fig:cont-G24}). 
The number of cores in each sub-clump varies from zero to more than ten (the maximum is 23). 
The estimated physical parameters and the number of gravitationally bound cores are summarized in Table~\ref{tab:subcl_phy}. 

\begin{deluxetable*}{lcccccccc}
\label{tab:subcl_phy}
\tabletypesize{\footnotesize}
\tablecaption{Sub-clump Physical Parameters}
\tablewidth{0pt}
\tablehead{
\colhead{Clump Name} & \colhead{Sub-clump Name} & \colhead{$M_\mathrm{sub-cl}$} & \colhead{$R_\mathrm{sub-cl}$} & \colhead{$\Sigma$} & \colhead{$n(\mathrm{H_2})$} & \colhead{$\lambda^\mathrm{th}_\mathrm{J, sub-cl}$} &\colhead{$M^\mathrm{th}_\mathrm{J, sub-cl}$} & \colhead{$N(\mathrm{Bound core})$} \\ 
\colhead{} & \colhead{} & \colhead{($M_\odot$)} & \colhead{(pc)} & \colhead{(g\,cm$^{-2}$)} & \colhead{($\times10^5$\,cm$^{-3}$)} & \colhead{(pc)} & \colhead{($M_\odot$)} & \colhead{}} 
    \startdata
    G010.991--00.082&1&15.0&0.064&0.24&4.45&0.048&0.76&3\\
    G010.991--00.082&2&9.0&0.041&0.35&2.56&0.032&0.5&0\\
    G010.991--00.082&3&25.3&0.07&0.34&1.95&0.041&0.66&8\\
    G014.492--00.139&1&69.4&0.059&1.33&4.45&0.02&0.35&5\\
    G014.492--00.139&2&26.2&0.039&1.12&2.56&0.018&0.31&3\\
    G014.492--00.139&3&16.4&0.043&0.60&1.95&0.026&0.45&0\\\enddata
\tablecomments{The radius is calculated from the geometric mean of the FWHM divided by 2. (This table is available in its entirety in machine-readable form.)}
\end{deluxetable*} 

\subsection{Jeans Length and Jeans Mass} 
\label{sec:Jeans_est} 
We estimated the thermal Jeans length and mass of clumps and sub-slumps from Equations~\ref{equ:Jlen} and \ref{equ:Jmass} with their density and dust temperature. 
The clump density is estimated from the flux density of the continuum emission obtained by the Atacama Pathfinder Experiment Telescope Large Area Survey of the Galaxy \citep[ATLASGAL; ][]{Schuller09}, and summarized in Table~1 of \citet{Morii23}. One example of the continuum image is shown in the left panel of Figure~\ref{fig:cont-G24}. 
The derived $\lambda^\mathrm{th}_\mathrm{J, cl}$ ranges from 0.06 pc to 0.22 pc, with a median of 0.12 pc, and $M^\mathrm{th}_\mathrm{J, cl}$ ranges from 1.1\,\Msun\,to 5.5\,\Msun, with a median of 2.4\,\Msun. 
Using the sub-clump density in Table~\ref{tab:subcl_phy}, sub-clumps' thermal Jeans length ($\lambda^\mathrm{th}_\mathrm{J, sub-cl}$) and mass ($M^\mathrm{th}_\mathrm{J, sub-cl}$) can be derived. 
The density of sub-clumps is about one order of magnitude larger than clumps, and the estimated $\lambda^\mathrm{th}_\mathrm{J, sub-cl}$ is a few times smaller than $\lambda^\mathrm{th}_\mathrm{J, cl}$, ranging from 0.018\,pc to 0.12\,pc, with a median of 0.05\,pc. $M^\mathrm{th}_\mathrm{J, sub-cl}$ ranges from 0.3\,$M_\odot$ to 3.5\,$M_\odot$, with a median of 1.1\,$M_\odot$. 

We also estimate the turbulent Jeans parameters of clumps. 
We applied a 1D Gaussian fitting toward the line profile of C$^{18}$O $J$ = 2--1 averaged within the clump, which was observed by TP. 
The obtained velocity dispersion ($\sigma_{\rm obs}$) is summarized in Table~\ref{tab:cl_Jeans_para}. 
The non-thermal velocity dispersion is given by $\sigma^2_\mathrm{nt} = \sigma^2_\mathrm{obs} - \frac{kT}{m_\mathrm{obs}}$ with $m_\mathrm{obs} = 28 m_\mathrm{H}$. By replacing $c_s$ with $\sqrt{c^2_s + \sigma^2_\mathrm{nt}}$ in Equation~\ref{equ:Jlen}, the turbulent Jeans length ($\lambda^\mathrm{tu}_\mathrm{J, cl}$) can be estimated. 
It is about five times larger than the thermal Jeans length ranging from 0.31 pc to 2.35 pc, and the turbulent Jeans masses ($M^\mathrm{tu}_\mathrm{J, cl}$) are estimated in a range of 10$^2$ to 7$\times$10$^3$ \Msun. 
All estimated values for each clump are summarized in Table~\ref{tab:cl_Jeans_para}. 

\begin{figure*}
    \centering
    \includegraphics[width=16cm]{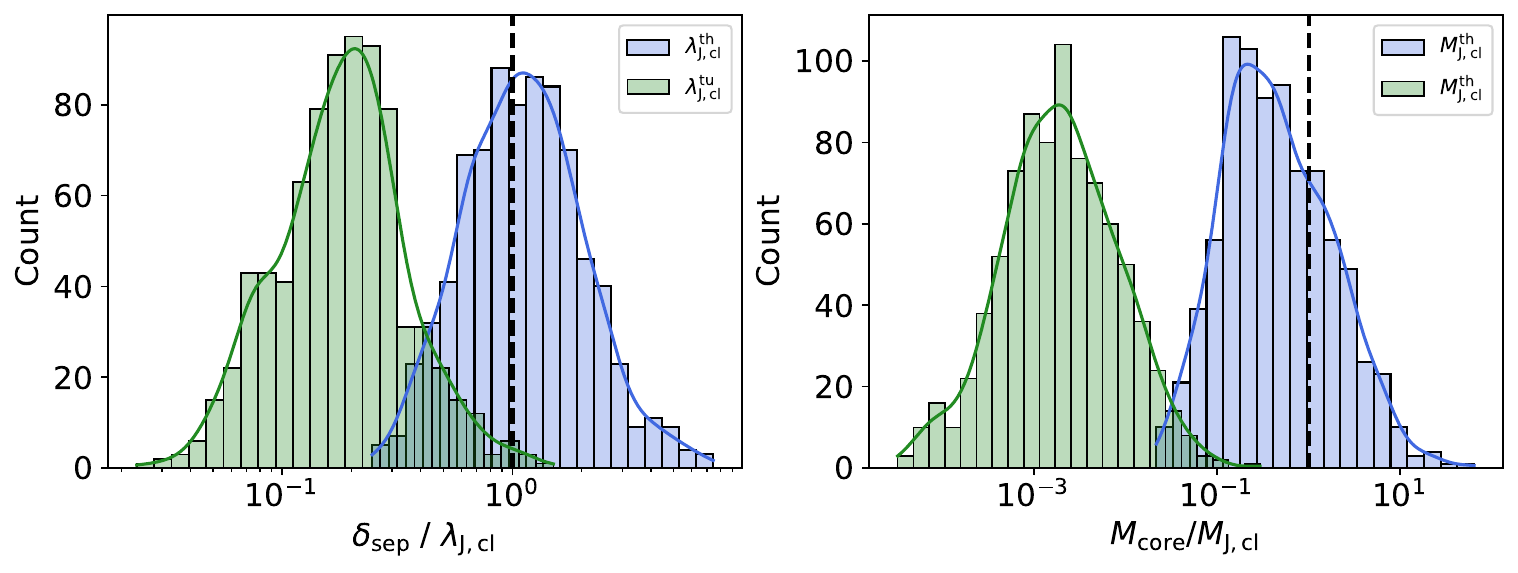}
    \caption{Core separation ($\delta_{\rm sep}$) and core masses normalized with Jeans length and Jeans mass of clumps, respectively. Blue and green bins represent the case of thermal or turbulent Jeans fragmentation, respectively. } The thick solid lines represent kernel density distribution. The vertical lines correspond to the ratio of unity. 
    \label{fig:sep_M_loghist}
\end{figure*}

Figure~\ref{fig:sep_M_loghist} shows the histogram of the ratios of the observed separation and mass divided by Jeans length and mass of clumps, respectively, in logarithmic scale.  
The top two panels show the case for the thermal Jeans fragmentation ($\delta_{\rm sep}/\lambda^\mathrm{th}_\mathrm{J, cl}$ and $M_{\rm core}/M^\mathrm{th}_\mathrm{J, cl}$), while the bottom two panels are the cases of turbulent Jeans fragmentation ($\lambda^\mathrm{tu}_\mathrm{J, cl}$ and $M^\mathrm{tu}_\mathrm{J, cl}$).  
The $\delta_{\rm sep}/\lambda^\mathrm{th}_\mathrm{J, cl}$ distributions take a peak around 1--2. For mass, the distribution has a large variation across more than two orders of magnitudes and more than half of cores have lower mass than the thermal Jeans mass of clumps. 
The peak is around 0.2 but there is a secondary peak around unity. 
For the turbulent Jeans fragmentation, both ratios of $\delta_{\rm sep}/\lambda^\mathrm{tu}_\mathrm{J, cl}$ and $M_{\rm core}/M^\mathrm{th}_\mathrm{J, cl}$ show their peaks at values much smaller than unity.  

\begin{deluxetable*}{lccccccccc}
\label{tab:cl_Jeans_para}
\tabletypesize{\footnotesize}
\tablecaption{Clump Information and Jeans Parameters}
\tablewidth{0pt}
\tablehead{
\colhead{Clump Name}  & \colhead{rms Noise\tablenotemark{a}} & 
\colhead{$\lambda^\mathrm{th}_\mathrm{J, cl}$} & \colhead{$M^\mathrm{th}_\mathrm{J, cl}$} 
& \colhead{$\sigma$\tablenotemark{b}} & 
\colhead{$\lambda^\mathrm{tu}_\mathrm{J, cl}$} & \colhead{$M^\mathrm{tu}_\mathrm{J, cl}$} & \colhead{$n({\rm Bound\,core})$\tablenotemark{c}} 
& \colhead{f(proto)\tablenotemark{d}}&\colhead{Q\tablenotemark{e}} \\ 
\colhead{} & \colhead{(mJy beam$^{-1}$)} & \colhead{(pc)}  & \colhead{($M_\odot$)} & \colhead{(km\,s$^{-1}$)} & \colhead{(pc)}  & \colhead{($M_\odot$)} & \colhead{} & \colhead{} & \colhead{}} 
    \startdata
    G010.991-00.082&0.88&0.08&1.3&1.1&0.46&230&14&0.29&0.78\\
    G014.492-00.139&1.90&0.06&1.0&1.8&0.53&660&21&0.71&0.81\\
    G015.203-00.441&0.90&0.11&2.9&1.0&0.43&170&--&--&0.78\\
    G016.974-00.222&0.33&0.11&1.8&1.2&0.64&370&3&0.67&0.77\\
    G018.801-00.297&0.44&0.11&1.9&1.3&0.68&460&--&--&0.68\\
    G018.931-00.029&0.59&0.19&5.3&1.5&1.07&920&1&0.0&0.79\\
    G022.253+00.032&0.31&0.11&2.1&1.0&0.51&190&1&1.0&0.86\\
    G022.692-00.452&0.46&0.19&4.7&1.2&0.97&570&4&0.75&0.77\\
    G023.477+00.114&1.00&0.08&1.4&1.3&0.46&300&7&0.75&0.71\\
    G024.010+00.489&0.65&0.06&1.1&1.0&0.31&130&--&--&0.78\\
    G024.524-00.139&0.53&0.13&2.2&1.5&0.90&780&9&0.91&0.73\\
    G025.163-00.304&0.58&0.11&1.9&1.2&0.67&400&12&0.57&0.67\\
    G028.273-00.167&0.70&0.08&1.2&1.6&0.68&650&8&0.23&0.66\\
    G028.541-00.237&0.50&0.13&2.5&1.4&0.88&700&8&0.78&0.72\\
    G028.564-00.236&1.20&0.07&1.3&1.9&0.69&980&--&--&0.81\\
    G028.927+00.394&0.40&0.15&3.1&1.0&0.67&270&--&--&0.87\\
    G030.704+00.104&0.47&0.18&3.6&1.5&1.17&1000&--&--&0.73\\
    G030.913+00.719&0.40&0.09&1.4&0.9&0.37&110&6&0.67&0.91\\
    G033.331-00.531&0.26&0.21&4.0&2.0&1.84&2740&--&--&0.68\\
    G034.133+00.076&0.41&0.19&4.1&1.2&1.04&640&4&0.4&0.81\\
    G034.169+00.089&0.32&0.19&4.4&1.0&0.76&310&2&0.5&0.77\\
    G034.739-00.119&0.42&0.10&1.7&1.2&0.60&360&17&0.63&0.75\\
    G036.666-00.114&0.39&0.10&1.8&0.9&0.43&160&7&0.71&0.82\\
    G305.794-00.096&0.61&0.08&1.7&1.3&0.45&300&21&0.5&0.78\\
    G327.116-00.294&0.71&0.11&2.2&1.4&0.70&520&2&0.75&0.71\\
    G331.372-00.116&0.73&0.16&2.9&1.8&1.31&1670&3&0.14&0.71\\
    G332.969-00.029&0.45&0.15&2.6&1.1&0.86&450&3&0.2&0.67\\
    G333.016-00.751&0.55&0.17&4.1&2.1&1.49&2570&1&0.0&0.83\\
    G333.481-00.224&0.65&0.13&3.3&1.2&0.65&400&12&0.36&0.76\\
    G333.524-00.269&1.4&0.08&2.4&1.5&0.49&450&19&0.77&0.84\\
    G337.342-00.119&0.36&0.19&3.7&2.8&2.35&6920&1&0.0&0.75\\
    G337.541-00.082&0.82&0.09&1.4&1.1&0.48&230&11&0.46&0.71\\
    G340.179-00.242&0.42&0.18&3.4&1.9&1.56&2190&0&0.0&0.82\\
    G340.222-00.167&0.75&0.11&2.3&1.0&0.52&220&5&0.0&0.81\\
    G340.232-00.146&1.1&0.14&2.5&2.1&1.28&2170&5&0.4&0.7\\
    G340.398-00.396&0.59&0.13&2.4&1.8&1.08&1330&10&0.1&0.79\\
    G341.039-00.114&0.72&0.13&2.5&1.1&0.63&290&16&0.37&0.82\\
    G343.489-00.416&0.60&0.10&1.3&1.0&0.50&190&--&--&0.71\\
    G345.114-00.199&0.50&0.08&1.1&1.1&0.43&210&--&--&0.75\\
    \enddata
    \tablenotetext{a}{The rms noise level of 7 m-array continuum image used for identifying sub-clumps. }
    \tablenotetext{b}{Velocity dispersion ($\sigma$) was obtained by the fitting of the line profile of C$^{18}$O ($J=2-1$) averaged within the clump with a 1D Gaussian, which was observed by Total Power (TP).}
    \tablenotetext{c}{Number of bound cores within $r=0.45$\,pc (see Section~\ref{sec:ncore}).}
    \tablenotetext{d}{The fraction of protostellar core to all bound cores in each clump.
    \tablenotetext{e}{The parameter to describe how centrally concentrated the core spatial distribution is. See \ref{sec:frag_div}, for more details. }}
\end{deluxetable*} 

\section{Discussion}
\label{sec:discuss}
In this section, we aim to characterize the fragmentation at early stages of high-mass star formation found in the ASHES sample by comparing core separation and mass with thermal and turbulent Jeans parameters, and by comparing clump and sub-clump Jeans parameters to study hierarchical fragmentation. 

\subsection{Thermal versus Turbulent Jeans Fragmentation}
\begin{figure*}
    \centering
    \includegraphics[width=14cm]{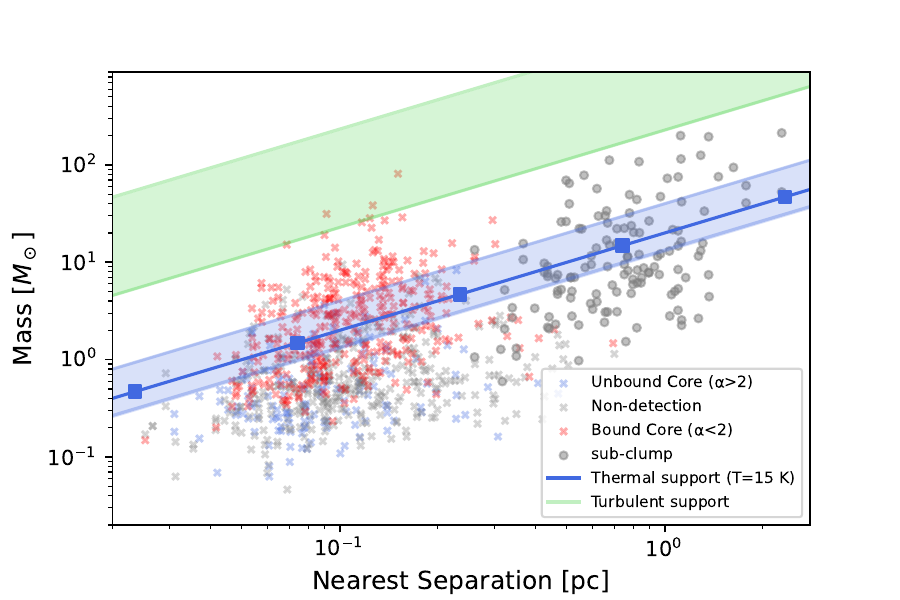}
    \caption{The relation between mass and nearest separation distance. The crosses represent cores identified in the ASHES survey \citep{Morii23}, and the gray circles correspond to sub-clumps. 
    The three different colors of the crosses indicate the gravitational states of cores; bound cores (red), unbound cores (blue), and cores without detections of dense gas tracers such as N$_2$D$^+$ and DCO$^+$ (gray). 
    The blue line shows thermal Jeans fragmentation with $T$ = 15 K and $n(\mathrm{H_2})$ = [10$^2$, 10$^6$] cm$^{-3}$, and the blue shaded region corresponds to the same density range but with T = [10, 30] K. 
    The five squares correspond to $n(\mathrm{H_2})$ of 10$^2$, 10$^3$, 10$^4$, 10$^5$, and 10$^6$\,cm$^{-3}$ from right to left. 
    The green-shaded region shows turbulent Jeans fragmentation to the same density and temperature range but $\sigma$ = [0.8, 2.7] km s$^{-1}$.}
    \label{fig:M-nns}
\end{figure*}

What dominates the fragmentation is one key question in star formation that has been investigated in some previous studies \citep[e.g.,][]{Zhang09, Palau13, Palau15, Sanhueza17,Sanhueza19, Beuther18_CORE, 
Beuther24}. 
We have the largest sample in IRDCs thanks to the mosaicked high spatial resolution and high sensitivity ALMA observations. IRDCs are thought to be the best for answering this question in the very early phase of high-mass star and cluster formation.  

To visualize that clearly, Figure~\ref{fig:M-nns} displays the mass--nearest separation relation, as in \citet[][ following this work, for this figure, we use the nearest separation between cores rather than the minimum separation defined in MST]{Wang14}. In this figure, blue and green shaded regions show what is expected from thermal Jeans fragmentation and turbulent Jeans fragmentation, respectively. 
Our identified cores and sub-clumps are denoted as crosses and circles, respectively. 
The blue line shows thermal Jeans fragmentation with $T$ = 15 K (mean temperature of the ASHES sample) varying the density from 10$^2$ cm$^{-3}$ to 10$^6$ cm$^{-3}$. The blue-shaded region shows the same density range but with $T$ from 10 K to 30 K. The green-shaded region shows turbulent Jeans fragmentation to the same density and temperature range but varying the velocity dispersion from 0.8 to 2.7 km s$^{-1}$. These ranges cover the ASHES sample properties \citep[see Table 1 of][]{Morii23}. 
The majority of cores (99.6\,\%) and all sub-clumps are plotted below the green area, and on average, both cores and sub-clumps are in the thermal Jeans fragmentation regime, while the variation of core masses is large, about two orders of magnitude. 
Especially gravitationally bound cores (red crosses) are located around blue-shaded regions, preferring thermal Jeans fragmentation. 

This trend can also be seen in Figure~\ref{fig:sep_M_loghist}.  
The $\delta_{\rm sep}/\lambda^\mathrm{th}_\mathrm{J, cl}$ distribution takes a peak around unity,  
 and both the mean and median values of $M_{\rm core}$/$M^\mathrm{th}_\mathrm{J, cl}$ are around unity (1.28 and 0.393, respectively). 
On the contrary, the Jeans length and mass in the turbulence-dominated case are both much larger than $\lambda^\mathrm{th}_\mathrm{J, cl}$ and $M^\mathrm{th}_\mathrm{J, cl}$, and the ratios become much smaller than unity (see the bottom panels).  
Thus, based on the estimated masses and separations, we conclude that core formation found in the ASHES sample (and likely in IRDCs in general) is regulated by thermal Jeans fragmentation rather than turbulent fragmentation. 

This conclusion, based on a large ALMA sample uniformly analyzed, consolidate earlier findings in IRDCs and active high-mass star-forming regions \citep[e.g.,][]{Palau13, Palau15, Liu17, Beuther15, Beuther18_CORE, Svoboda19, Sanhueza19, Liu19, Lu20, Beuther21, Saha22}. 
However, this result disagrees with some observations that suggest the importance of turbulent or magnetic field support \citep[e.g.,][]{Zhang09, Wang11, Pillai11, Zhang11, Wang14, Zhang15, Li19a, Rebolledo20}. 
One significant difference is the image fidelity reached by our observations (i.e., higher spatial resolution, mass sensitivity, and better UV coverage). 
For example, we revealed further fragments and succeeded in identifying about three times more cores than previous studies using SMA or Plateau de Bure Interferometer (PdBI), such as \citet[][G023.477+00.114]{Beuther13}, \citet[][G028.564-00.236]{Lu15}, \citet[][G028.273-00.167]{Sanhueza17}, \citet[][G014.492-00.139]{Li19a}, and \citet[][G010.991–00.082]{Pillai19}.  
Additionally, our mosaicked observations covered a larger FOV than in single-pointing observations. 
That is why we succeeded in resolving cores and identifying more cores with lower masses than in previous studies. 

Some other ALMA observations with similar setups like the ASHES survey still suggest turbulence-dominated fragmentation \citep[e.g.,][]{Rebolledo20, Xu23}, but most are not in a quiet/early stage. They focus on more evolved, active high-mass star-forming regions where some feedback from the newborn stars can suppress the fragmentation by inducing additional turbulence and warming up the surrounding gas \citep[e.g.,][]{KrumholzMcKee08}. 

\subsection{Hierarchical fragmentation}
We found that the overall properties of identified cores are consistent with thermal Jeans fragmentation as shown in Figure~\ref{fig:M-nns}. 
However, as Figure~\ref{fig:cont-G24} shows, cores are embedded in sub-clump structures inside clumps. 
We next address if cores form from the fragmentation of clumps or dense sub-structures (here we refer to them as sub-clumps) by using clumps' and sub-clumps' Jeans parameters. 
In this section, we use only gravitationally bound cores which will likely form stars. Core separations are recalculated only using bound core positions. 

\begin{figure*}
    \centering
    \includegraphics[width=15cm]{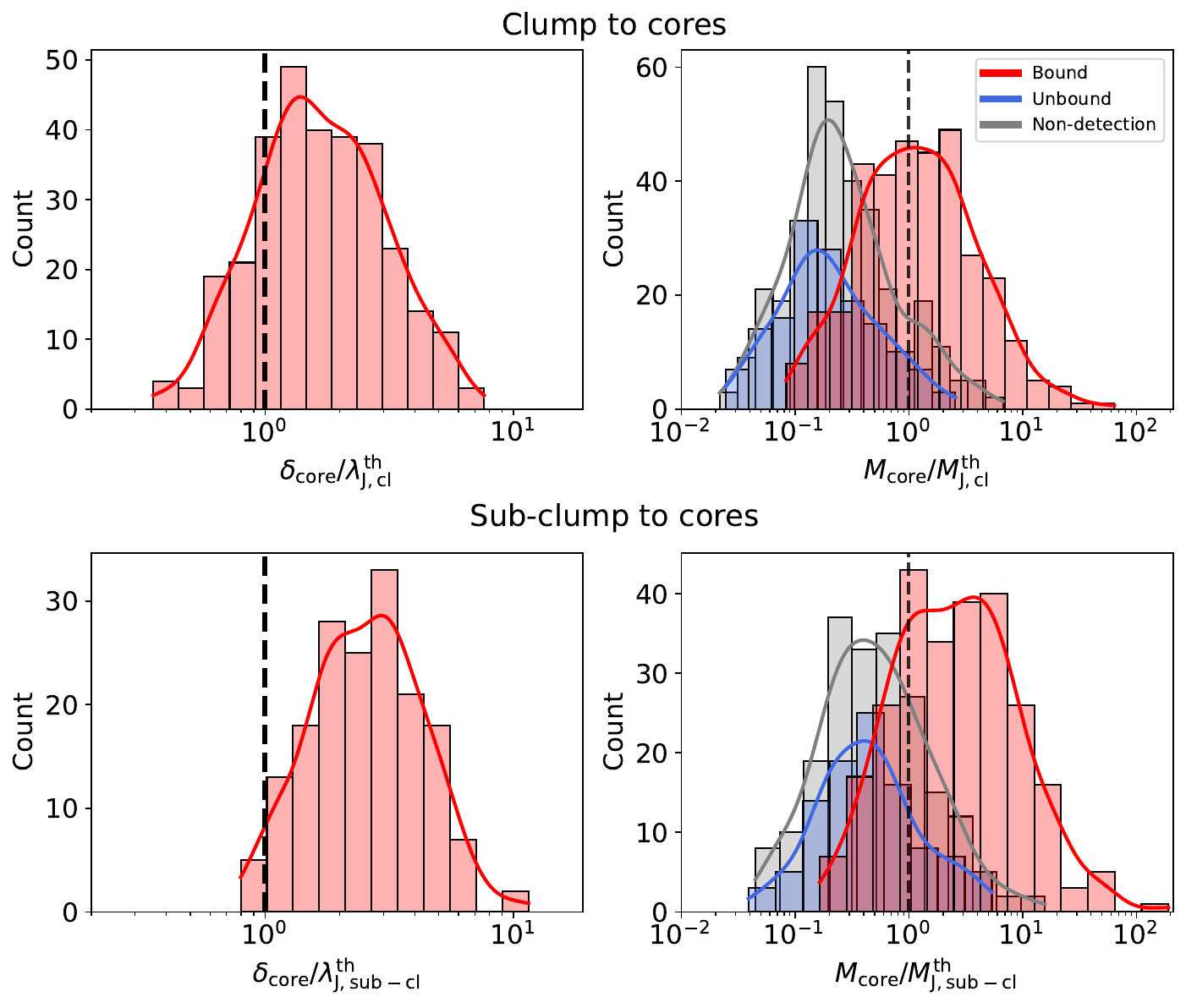}
    \caption{Core separation ($\delta_{\rm sep}$) and core masses normalized with thermal Jeans length and masses, respectively. The top three panels show ratios normalized by thermal Jeans lengths and masses of clumps, and the bottom three show the case of sub-clump fragmentation. Only cores inside sub-clumps are used for this analysis. The thick solid lines represent kernel density distribution. The vertical lines correspond to the ratio of unity. (Left) Core separations of bound cores normalized by Jeans length. (Right) Core masses normalized by Jeans mass. Gravitationally bound and unbound are colored red and blue, respectively (cores without detections of dense gas tracers are in gray). 
    }
    \label{fig:hist-cl-subcl}
\end{figure*}
Figure~\ref{fig:hist-cl-subcl} compares the fragmentation of clumps and sub-clumps with the observed core properties. 
The top two panels show the ratios of separation and masses normalized by thermal Jeans parameters estimated by using clump density, and the bottom two panels show the case of using sub-clump density.  For the bottom panels, only cores inside sub-clumps are considered. 
The left panels present the ratios of separation of bound cores normalized by thermal Jeans length. Their distribution peaks are located around 1--2 and 2--3 for the top and bottom, respectively. 
The right panels display the histogram of the mass ratios and bound cores are highlighted in red. 
The peak is just around unity for clump fragmentation. Although the bottom panel shows no clear single peak, it has a broad peak around 1--5. 
These histograms also show that unbound cores and cores without the detection of dense gas tracers generally have masses smaller than the Jeans mass of clumps. 
Overall, the observed core properties favor the fragmentation from clumps rather than sub-clumps. 

Hierarchical fragmentation is expected in one of the theoretical scenarios for high-mass star formation, called Global Hierarchical Collapse (GHC) scenario \citep[][]{vazquez19}. 
Such hierarchical fragmentation has been reported in some previous studies observing high-mass starless clump candidates \citep[][]{Svoboda19, Zhang21}, IRDCs \citep[][]{Wang11, Wang14}, and also OMC-1S \citep[][]{Palau18}.  
Our analysis also revealed sub-clumps, intermediate structures connecting clumps and cores, but the core properties cannot be explained well by sub-clump fragmentation. This implies that a step-by-step fragmentation (clump to sub-clump to core) is unlikely.

We also investigated the fragmentation from clump to sub-clump by comparing Jeans parameters with sub-clump separations and masses. We find no clear peak in the distributions due to the small number of statistics. However, thermal Jeans fragmentation may still be favored over turbulent fragmentation as shown in Figure~\ref{fig:M-nns}. We note that the sub-clumps locate, in Figure~\ref{fig:M-nns}, in the area at which the parental clumps should have densities of $\sim$10$^3$ cm$^{-3}$, inconsistent with observations. This could be explained if sub-clump structures form as a result of clump fragmentation into cores. Sub-clumps could be the ensemble of cores that evolve as cores evolve, not in a step-by-step hierarchical fragmentation, but rather in a simultaneous formation process. 

We point out that this simultaneous formation of sub-clumps and cores, in which the core properties are not determined by sub-clumps, may only be valid at the very early stages of high-mass star formation traced in the ASHES survey. Later, in more active high-mass star-forming regions, once subsequent fragmentation occurs, core properties may be explained differently. 

\subsection{Fragmentation Level and Clump Properties} 
\label{sec:ncore}
\begin{figure*}
    \centering
    \includegraphics[width=18cm]{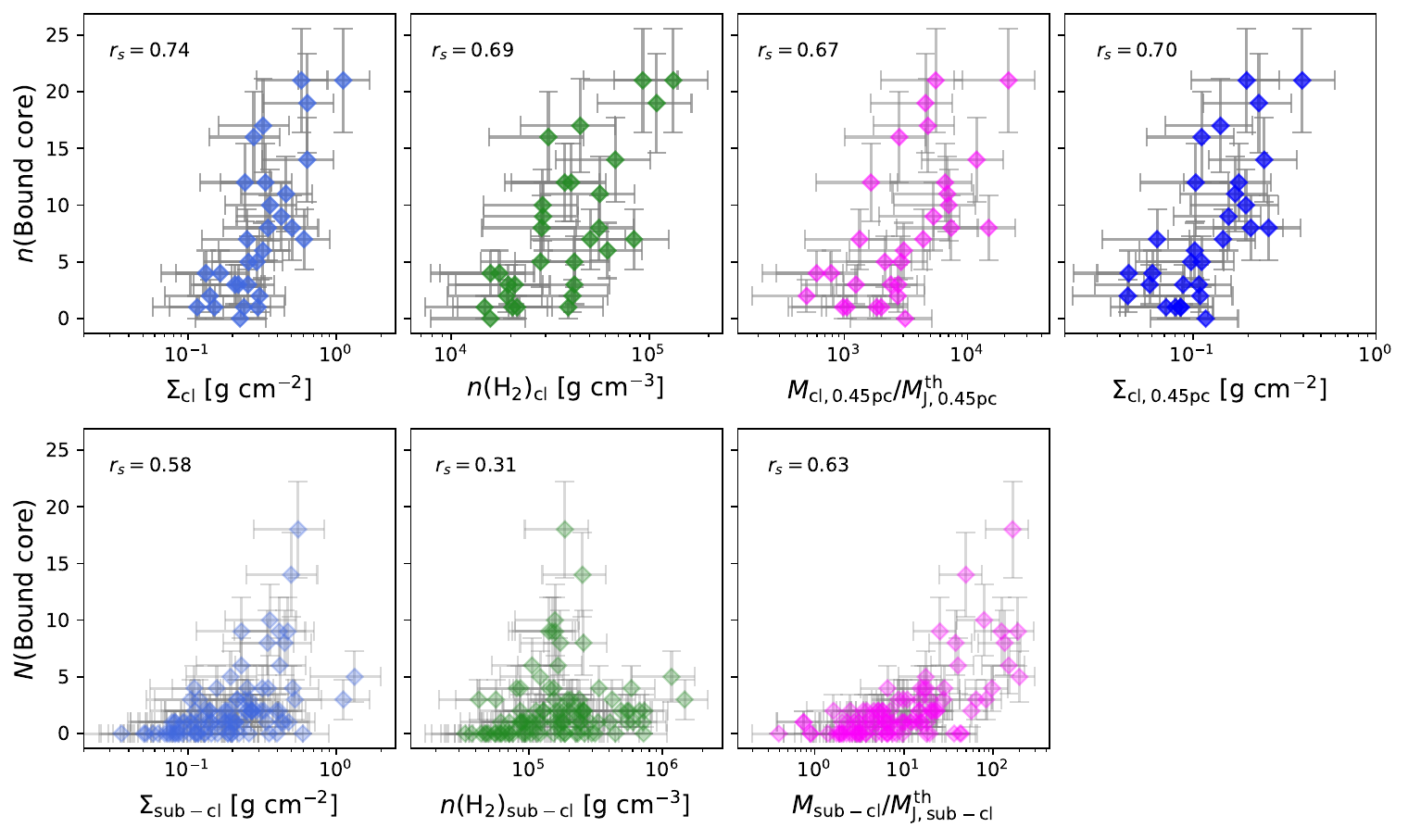}
    \caption{(Top) The number of bound cores above 0.41\,$M_\odot$ located around the clump peak ($r<0.45$\,pc), $n(\mathrm{Bound\,core})$, as a function of clump surface density $\Sigma_\mathrm{cl}$, clump mean number density $n({\rm H_2})_{\rm cl}$, Jeans number, the ratio of $M_\mathrm{cl, 0.45 pc}$ to thermal Jeans mass ($M^\mathrm{th}_\mathrm{J, 0.45 pc}$), and clump surface density with a limited area $\Sigma_\mathrm{cl, 0.45 pc}$for selected 30 clumps. 
    Spearman’s rank correlation coefficients are denoted inside each panel. All p--values are much smaller than 0.01. 
    (Bottom) The number of bound cores above 0.41\,$M_\odot$ in sub-clumps, $N(\mathrm{Bound\,core})$, as a function of sub-clump surface density $\Sigma_\mathrm{sub-cl}$, mean number density $n({\rm H_2})_{\rm sub-cl}$, and Jeans number ($M_\mathrm{sub-cl}/M^\mathrm{th}_\mathrm{J, sub-cl}$).}
    \label{fig:ncore_sigma}
\end{figure*} 
Jeans fragmentation invokes the idea that the fragmentation level or the number of cores depends on clump density. 
We investigated the correlation of clump density or Jeans number (the ratio of clump mass to Jeans mass) with the number of bound cores, $n(\mathrm{Bound\,core})$. 
To calculate $n(\mathrm{Bound\,core})$, we count the number of bound cores within the same physical area for all clumps and impose a mass threshold to reduce the effect of having different distances and sensitivities for each clump. 
We have limited the sample for this discussion, excluding clumps that are located too close ($< 3.5$\,kpc) and too far ($> 5.5$\,kpc), and two more with the worst mass sensitivity ($> 0.45 M_\odot$). 
As a result, the 30 clumps remaining are located between  3.5 and 5.5 kpc and have a mass sensitivity between 0.086 and 0.41 $M_\odot$. 
We count cores within a circle with a radius of 0.45\,pc centered on the mean positions of cores. The circle size almost corresponds to the FoV of the closest clump. 
In addition, we impose for all clumps a mass threshold of 0.41 $M_\odot$, which corresponds to the worst mass sensitivity among the 30 clumps. 
The measured $n(\mathrm{Bound\,core})$ are listed in Table~\ref{tab:cl_Jeans_para}). 

The top-left two panels of Figure~\ref{fig:ncore_sigma} show moderate to strong correlations of $n(\mathrm{Bound\,core})$ with surface density $\Sigma_{\rm cl}$ and mean clump number density $n(\mathrm{H_2})_{\rm cl}$. 
It indicates denser clumps produce more number of cores (higher fragmentation level). 
We found stronger correlation between $n(\mathrm{Bound\,core})$ and $\Sigma_\mathrm{cl}$ with a Spearman’s rank correlation coefficient of $r_s = 0.74$, while that is $r_s = 0.69$ for $n_\mathrm{cl}$. Both p-values are much smaller than 0.01.  
The less scattered plot of $n(\mathrm{Bound\,core})$--$\Sigma_\mathrm{cl}$ implies that the clump surface density is the best indicator of the fragmentation level: the higher the clump surface density, the more likely it is to have a larger number of cores. 
To confirm that these correlations do not result from the co-dependence on the distance, we re-calculated clump mass ($M_\mathrm{cl, 0.45 pc}$) and surface density ($\Sigma_\mathrm{cl, 0.45 pc}$) using flux within the same physical area (r=0.45 pc). 
The right panel displays the strong correlation between $\Sigma_\mathrm{cl, 0.45 pc}$ and $n(\mathrm{Bound\,core})$. 
We also compared $n(\mathrm{Bound\,core})$ with the Jeans numbers ($M_\mathrm{cl, 0.45\,pc} / M^\mathrm{th}_\mathrm{J, 0.45\,pc}$), which is proportional to $\Sigma^3_\mathrm{cl, 0.45\,pc} / n(\mathrm{H_2})^{3/2}_\mathrm{cl, 0.45\,pc}$. 
It is confirmed that the measured $n(\mathrm{Bound\,core})$ has a strong correlation with the number of cores expected from thermal Jeans fragmentation ( $r_s = 0.70$ and p-value$<< 0.01$). 

These correlations are still found in the case of sub-clump to core fragmentation, but in this case, the coefficients are relatively weaker. 
The bottom three panels also imply the strong or moderate correlation between the number of bound cores in each sub-clump and sub-clump surface density, volume density, and Jeans number, although relatively weaker than the clump case. It suggests that sub-clumps are also involved with core formation. 

To summarize, we revealed that a higher fragmentation level (or higher core number density) can be expected from a region with a higher surface density and it is consistent with Jeans fragmentation.  
The tight correlation between the fragmentation level and the clump/cloud surface density has been observed in more evolved star-forming regions as well, indicating that this correlation begins early on in IRDCs and prevails during the evolution of high-mass star-forming regions \citep[e.g.,][]{Palau14, Sokol19}. 

\subsection{Fragmentation Diversity}\label{sec:frag_div}
The ASHES survey also reveals a diversity in both the range of core masses per clump (mass dynamic range) and fragmentation patterns.  

\begin{figure*}
    \centering
    \includegraphics[width=14cm]{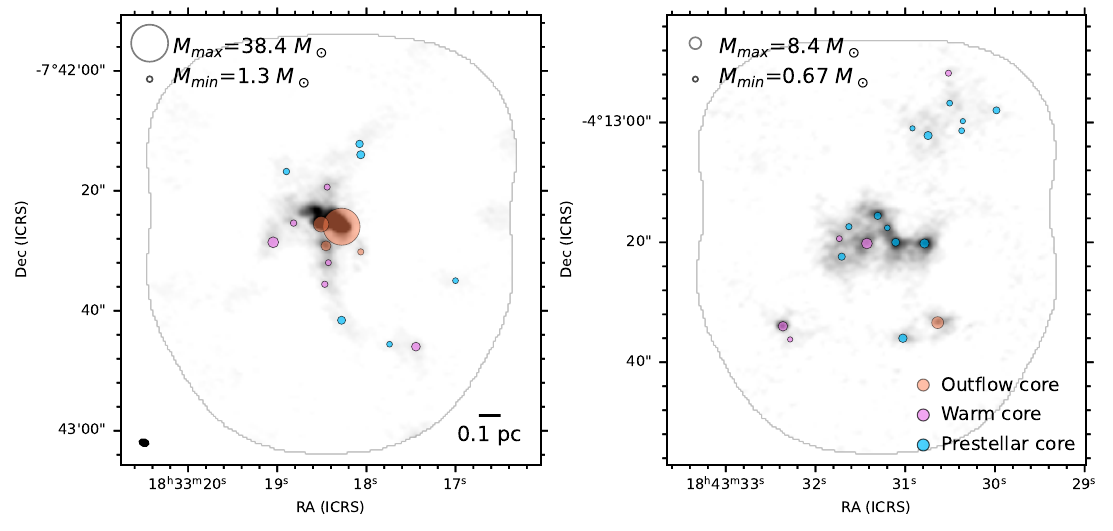}
    \caption{ALMA 1.3 mm continuum image of (left) G024.010+00.489  and (right) G028.273--00.167. The circle size represents the core mass, and the position is centered at the continuum peak of each core. The three different colors have meanings the same as Figure~\ref{fig:M-nns}. }
    \label{fig:cont_massdyn}
\end{figure*} 
\citet{Morii23} reported that most clumps host low- to intermediate-mass cores. However, the dynamic range in core masses varies from clump to clump. 
For example, the left panel in Figure~\ref{fig:cont_massdyn} shows some (relatively) massive cores surrounded by some low-mass cores, and the right panel shows a cluster of low-mass cores with a small dynamic range in mass. 
\begin{figure*}
    \centering
    \includegraphics[width=10cm]{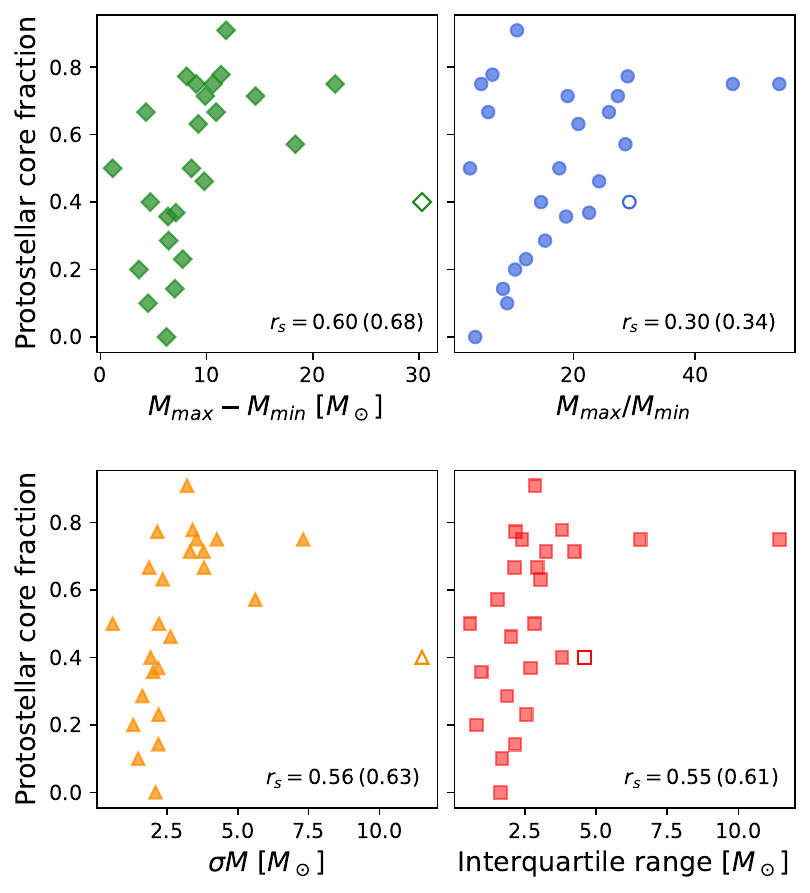}
    \caption{Correlation plots between the fraction of protostellar core and the mass dynamic range (the difference and the ratio of the most massive core to the least massive core, the standard deviation, and the interquartile range) for each clump. Additional five clumps with less than one bound core are excluded from the sample in Figure~\ref{fig:ncore_sigma}. The unfilled marker corresponds to G340.232--00.146 (see the main text). Spearman’s rank correlation coefficients are given within each panel, and the values in parentheses are the case that the unfilled point is excluded. P-values are all less than 0.01 except the top-right panel (0.1). }  
    \label{fig:fproto}
\end{figure*}
We find that such a mass dynamic range correlates with the fraction of protostellar cores (cores with outflow or warm line detection). 
Figure~\ref{fig:fproto} shows the fraction of bound protostellar cores to the total number of bound cores as a function of the maximum and minimum mass difference ($M_{\rm max}-M_{\rm min}$), the ratio ($M_{\rm max}/M_{\rm min}$), the standard deviation of core masses ($\sigma M$), and interquartile range. 
The fraction of protostellar cores is summarized in Table~\ref{tab:cl_Jeans_para}. 
The x-axis of the top two panels is derived from the maximum and minimum core masses and can be significantly affected by a single peculiar object if exists. The interquartile range, the difference of 25 percentile and 75 percentile, is on the contrary, less affected by the most massive object. 
The Spearman’s rank correlation coefficients are $r_s \sim 0.5$ and p–values are less than 0.01 except for the case of $M_{\rm max}/M_{\rm min}$. 
One outlier in these plots is G340.232--00.146, shown as an unfilled marker. As discussed in \citet{Sanhueza19}, in this peculiar clump, the most massive core is rather large with a radius of $\sim$10$^4$\,au and more fragmented structures are expected in higher angular resolution observations from visual inspection of the continuum image. The correlation coefficients displayed in Figure~\ref{fig:fproto} become higher if we exclude it as shown in the values in parenthesis (bottom, right of each panel). 
These plots imply that if we take the fraction of protostellar cores as an indicator of cluster evolution, the dynamic range of mass increases with cluster evolution. 
This correlation can be interpreted as clump-fed accretion onto cores. Clumps would initially fragment into cores of similar mass (small dynamic mass range), and as time goes on, some cores grow in mass more than others, resulting in a larger mass range difference (large dynamic mass range).   

Among 39 ASHES targets, some clumps show aligned fragmentation as the left panel of Figure~\ref{fig:cont_massdyn}, some show concentrated, and some show spread core distributions with several sub-clumps (e.g., right panel of Figure~\ref{fig:cont_massdyn}). 
We calculated the $\mathcal{Q}$-parameter to investigate the cluster members' distribution. 
The $\mathcal{Q}$-parameter was defined by \citet{Cartwright_Whitworth04} as  
\begin{equation}
    \mathcal{Q} = \frac{\Bar{m}}{\Bar{s}}.
\end{equation} 
The term $\Bar{m}$ is the normalized mean edge length of MST and is defined as 
\begin{equation}
    \Bar{m} = \frac{\Sigma^{N_\mathrm{c}-1}_{i=1} L_i}{N_\mathrm{c}-1}\times \frac{N_\mathrm{c}}{(N_c A)^{1/2}} =  \frac{\Sigma^{N_\mathrm{c}}_{i=1} L_i}{(N_c A)^{1/2}},
    \label{equ:bar_m}
\end{equation}
where $N_\mathrm{c}$ is the number of cores in the region, $ \Sigma^{N_\mathrm{c}}_{i=1} L_i$ is the total length of all the lines MST connected, hereafter `edges', and $A$ is the area of the cluster and estimated by $A=\pi R^2_\mathrm{cluster}$. Here, the radius of the cluster $R_\mathrm{cluster}$ is defined as the distance from the mean position of cores to the farthest core position. 
The second term in the first equation of Equation~\ref{equ:bar_m} is the factor to normalize the mean edge length of cores (the first term is defined as $l_\mathrm{MST}$) having different areas ($A$) and/or different numbers of cores ($N_\mathrm{c}$). 
The term $\Bar{s}$ is the ratio of the mean core separation to the cluster radius ($R_\mathrm{cluster}$). 
Here, the core separation is different from the minimum core separation ($\delta_{\rm sep}$), and it is the core separation within the region, not only considering the minimum separation.  
Now both $\Bar{m}$ and $\Bar{s}$ are independent of the number of cores in the cluster-forming clump. 

For clusters with a smooth radial density gradient ($n\propto r^{-\alpha}$), $\mathcal{Q}$ increases from $\sim$0.8 to 1.5 as the degree of concentration increases from $\alpha =0$ to 2.9, and for sub-clustering clusters $\mathcal{Q}$ becomes smaller than 0.8 \citep{Cartwright_Whitworth04}.   
The parameters estimated for the ASHES sample range from 0.6 to 0.9, and their average is 0.76. 
Some clumps indicate a uniform distribution of cores ($\mathcal{Q} \sim$0.8), but most (70\%) prefer sub-clustering ($\mathcal{Q}<0.8$). 
Clumps with aligned fragmentation have a $\mathcal{Q}$-parameter of 0.7--0.8. However, we note that, as can be inferred from the definition of the $\mathcal{Q}$-parameter, this parameter is unable to identify aligned fragmentation. 
Clumps with several sub-clumps or showing spread fragmentation have $\mathcal{Q}<0.8$. 

The origin of such variation in the fragmentation pattern is not yet clear from the current data (e.g., clump mass, density, virial parameter, protostellar, or core fraction), and further information on the magnetic field or clump-scale properties such as large-scale gas dynamics seems to be necessary. 
For example, \citet{Tang19} suggests that the balance among the magnetic field, turbulence, and gravity determines the core fragmentation pattern such as no fragmentation, aligned fragmentation, and clustered fragmentation. 
It should be noted that we find no clear correlation between the dynamic range in mass and the fragmentation pattern; both aligned fragmentation and spread fragmentation show large and small mass dispersion. 

\subsection{Early Fragmentation picture}
We have revealed that the observed mean core separation and masses are comparable to thermal Jeans lengths and masses, respectively, and much smaller than turbulent Jeans parameters. 
It implies that turbulence is not a dominant source characterizing core formation. 
This is consistent with the study by \citet{Traficante20}, indicating that gravity dominates over turbulence once the regions become dense (e.g., $\Sigma > 0.1$\,g\,cm$^{-2}$). ASHES targets are all dense with a surface density larger than 0.1\,g\,cm$^{-2}$. 
Compared with other ALMA studies in high-mass star-forming regions, in which core properties are better explained by turbulent Jeans fragmentation \citep[e.g., ][]{Rebolledo20, Xu23}, the ASHES  sample contains 70\,$\mu$m-dark, cold regions not affected by feedback mechanism from massive stars. 
Thus, our finding implies that the initial fragmentation in massive clumps, prior to the changes due to gravitational accretion and some feedback effects, is described by thermal Jeans fragmentation. 

However, our sample still contains super-Jeans cores with a mass more than 10 times $M^\mathrm{th}_\mathrm{J, cl}$. 
Those cores may have grown in mass by acquiring gas from the surrounding environment. Infall rates in the range 10$^{-4}$--10$^{-3}$ have been measured in two ASHES targets \citep{Contreras18,Redaelli22}, allowing the cores to quickly grow in mass in a free-fall time. Alternatively, the magnetic field may play a role in suppressing fragmentation as suggested by theoretical studies \citep{Hennebelle08, Commercon11}.  
The theoretical prediction that the magnetic field can inhibit fragmentation in high-mass star-forming regions has garnered support from observational studies in more evolved high-mass star-forming regions.  
Dust polarization emission from infrared-bright sources has highlighted the significant role magnetic fields play in exerting pressure on the medium, from clump to core scale, effectively curbing fragmentation during gravitational collapse \citep{Zhang14, HullZhang19}. 
Supporting evidence for fragmentation suppression due to magnetic fields has also been presented by \citet{Frau14}. 
In addition, \citet{Das_Basu_Andre21} showcased the magnetic field’s impact on reducing fragment numbers, while \citet{Palau21} reported a tentative correlation between fragment quantity and the mass-to-flux ratio among massive dense cores, as suggested by theoretical and numerical works. 
Future observations of dust polarisation toward such sub-clumps or massive cores would verify this effect on suppressing fragmentation. 
Studying how significantly the magnetic field contributes to the fragmentation is also important to the understanding of the diversity of fragmentation patterns seen in Figure~\ref{fig:cont_massdyn}. Although at relatively smaller scales, these points are one of the goals of the MagMaR \citep[Magnetic Fields in Massive Star-forming Regions;][]{Fernandez21,Cortes21,Sanhueza21} survey once the whole survey sample is analyzed. 

We found sub-structures inside clumps using the 7\,m-array data, which is consistent with the measured $\mathcal{Q}$ of 0.6-0.8, implying that the initial core distribution is not yet so concentrated but rather sub-clustered. 
These sub-structures are located in the thermal Jeans fragmentation-dominated regime in Figure~\ref{fig:M-nns}, as well as cores. 
We therefore investigated whether cores are directly formed from such sub-clumps rather than from clumps. 
We find no evidence that sub-clump fragmentation is the preferred mechanism to explain the observed core properties. 
The comparison of the number of cores or the degree of fragmentation with the properties of clumps or sub-clumps also suggests a stronger link between clumps and cores. 
A possible picture is that sub-clumps and cores are simultaneously formed from clumps, and core properties are determined from clump properties. 

It should be noted that such sub-clumps are likely to contribute to density segregation since denser cores are generally embedded in sub-clumps. 
If they can grow by more effective gas feeding, this may later lead to mass segregation as discussed in \citet{Morii23}. Gas feeding or gravitational collapse of clumps and sub-clumps would increase the mass dynamic range as seen in Figure~\ref{fig:fproto}. 
\citet{Xu23-arxiv} suggests that the gravitational concentration or gas accretion towards the center of mass would result in the appearance of mass segregation and in the increase of the $\mathcal{Q}$-parameter. 
Comparing our results with more evolved clusters would confirm this hypothesis.

\section{Conclusions}
\label{sec:conclusion}
We have studied the fragmentation properties in 39 clumps as a part of the ALMA Survey of 70 $\mu$m dark High-mass clumps in Early Stages (ASHES), which aims to characterize the very early phase of high-mass star formation.  
Using the 839 cores identified in the continuum images, we compared their masses and separations with Jeans parameters. 
We have obtained the following conclusions:

\begin{enumerate}
    \item The mean core separation measured by the MST method ranges from 0.08 pc to 0.32 pc in each region, and core masses range from 0.05 \Msun to 81 \Msun. 
    The core mass and core separation are explained by thermal Jeans fragmentation ruling out turbulent Jeans fragmentation at the very early stages of high-mass star formation. 
    \item Comparing the Jeans parameters of clumps and sub-clumps with the observed core properties, core properties, especially for bound cores, are likely determined from clumps. We interpret this as a simultaneous formation of sub-clumps and cores within clumps. 
    \item The fragmentation level or the number of cores within each clump shows a strong correlation with the Jeans number, the ratio of clump mass to Jeans mass, implying that early core formation can be described with thermal Jeans fragmentation. It also has a strong correlation with clump surface density. 
    \item Furthermore, our sample shows the diversity of fragmentation in terms of mass dynamic range and spatial distribution. The correlation between the protostellar core fraction and the mass dynamic range is likely a sign of the clump-fed accretion scenarios. We have revealed aligned, spread, clustered, and sub-clustered fragmentation patterns, and the measured $\mathcal{Q}$-parameter also implies that the early fragmentation seen in ASHES fields is not centrally concentrated.  
\end{enumerate}

\begin{acknowledgments}
K.M. is financially supported by Grants-in-Aid for the Japan Society for the Promotion of Science (JSPS) Fellows (KAKENHI Number 22J21529) and supported by FoPM, WINGS Program, the University of Tokyo. K.M. is also supported by JSPS Overseas Challenge Program for Young Researchers (202280210). 
PS was partially supported by a Grant-in-Aid for Scientific Research (KAKENHI Number JP22H01271 and JP23H01221) of JSPS. 
GS acknowledges the projects PRIN-MUR 2020 MUR BEYOND-2p (``Astrochemistry beyond the second period elements'', Prot. 2020AFB3FX) and INAF-Minigrant 2023 TRIESTE (``TRacing the chemIcal hEritage of our originS: from proTostars to planEts''; PI: G. Sabatini). 
Data analysis was in part carried out on the Multi-wavelength Data Analysis System operated by the Astronomy Data Center (ADC), National Astronomical Observatory of Japan. 
This paper uses the following ALMA data: ADS/JAO.ALMA\#2015.1.01539.S, 
ADS/JAO.ALMA\#2017.1.00716.S, and ADS/JAO.ALMA\#2018.1.00192.S. 
ALMA is a partnership of ESO (representing its member states), NSF (USA) and NINS (Japan), together with NRC (Canada), $MOST$ and ASIAA (Taiwan), and KASI (Republic of Korea), in cooperation with the Republic of Chile. The Joint ALMA Observatory is operated by ESO, AUI/NRAO, and NAOJ. 
\facility {ALMA} 
\software{CASA (v4.5.3, 4.6, 4.7, 5.4, 5.6; \citealt[][]{CASA22})}
\end{acknowledgments}

\bibliography{reference}
\bibliographystyle{aasjournal}

\end{document}